\documentclass[twocolumn,pre,superscriptaddress,preprintnumbers,amsmath,amssymb,longbibliography]{revtex4-1}
\usepackage{graphicx}
\usepackage{color}
\usepackage{stackengine}

\def\gtrless{\raise2.5pt\hbox{$>$}\llap{\lower2.5pt\hbox{$<$}}}
\def\gtrapprox{\raise2.5pt\hbox{$>$}\llap{\lower2.5pt\hbox{$\approx$}}}

\newcommand{\bsq}[1]{\begin{subequations}\label{#1}}
\newcommand{\esq}{\end{subequations}}
\newcommand{\beq}[1]{\begin{equation}\label{#1}}
\newcommand{\eeq}{\end{equation}}
\newcommand{\beqa}[1]{\begin{eqnarray}\label{#1}}
\newcommand{\eeqa}{\end{eqnarray}}

\newcommand{\mb}{\mathbf}

\newcommand{\rb}{{\bf r}}

\newcommand{\vb}{{\bf v}}

\renewcommand{\rho}{\varrho}
\renewcommand{\epsilon}{\varepsilon}

\begin{document}

\title{Chaos and mixing in self-propelled droplets}

\author{Reiner Kree}
\author{Annette Zippelius}
\affiliation{University G\"ottingen, D-37077 G\"ottingen, Germany}

\date{\today}

\begin{abstract} We consider self-propelled droplets which are driven
  by internal flow. Tracer particles, which are advected by the flow,
  in general follow chaotic trajectories, even though the motion of the
  autonomous swimmer is completely regular. The flow is mixing, and
  for P\'{e}clet and Batchelor numbers, which are realized e.g. in eucaryotic
  cells, advective mixing can substantially accelerate and even dominate transport by diffusion.
  \end{abstract}

\pacs{82.70.Dd, 61.20.Ja}

\maketitle

\section{Introduction}
Fluid flow in small droplets or vesicles is essential for many
biological and chemical processes, both in artificially fabricated
microdroplets and in biological cells.  Droplet
microfluidics~\cite{Teh2008, Chou2015} has a rapidly growing range of
applications including molecular detection, imaging, drug delivery and
diagnostics. It has also been used to synthesize artificial cells,
using droplet stabilized vesicles which can be filled with filaments
and motors in a controlled way~\cite{Spatz2017}. Furthermore, viscous
liquid droplets, which are chemically driven out of equilibrium, have
been shown to grow and divide, reminiscent of living
cells~\cite{Zwicker2}.

 Both artificial and biological cells contain
active matter, which can generate intracellular flow, which in turn
leads to a variety of functions, ranging from transport of
nutrients~\cite{Goldstein2008,Goldstein2015} to control of asymmetric
cell division~\cite{Mittasch2018} and to cell
locomotion~\cite{Kree_2017}. 
The internal flow in droplets and cells is
experimentally accessible with the help of tracer particles which are
advected by the flow. Several techniques  have been used, such as video
microscopy, laser based particle tracking and FCS to just mention a
few,
and a variety of tracers are available~\cite{Theriot2009,Sackmann2008}.
Recently it has become possible to perturb cytoplasmic flow locally, in vivo and probe-free
by an interactive microscopic technique~\cite{Mittasch2018}, which allows to simultaneously observe the consequences of such perturbations.

Among the many possible functions of internal flow is the advection of particles, which gives rise to stirring and mixing as important
prerequisites for biochemical reactions.
In microdroplets, laminar
creeping flow often persists, and mixing by diffusion is inefficient
on reaction time scales.  In droplet-based microfluidics several
techniques have been developed to rapidly mix the particle content by an
externally generated flow.  Stirring by chaotic advection due to
simple Eulerian flow fields has been extensively studied in the past
(for a recent review see \cite{Aref_2017}). The first example of an
incompressible flow field inside a sphere, which leads to chaotic
Lagrangian trajectories was given in Ref.~\cite{Moffatt1990}. In
Ref. ~\cite{Stone1991} it was shown that simple linear flow fields in
the ambient fluid may generate stirring by chaotic advection in
passive droplets. 
In eukaryotic cells, one observes intracellular flow, known as cytoplasmic streaming, which is generated for example by molecular motors, carrying cargo and entraining the adjacent fluid. Advection of chemicals by the flow contributes significantly to molecular transport and mixing \cite{Goldstein2008,Goldstein2015}. The  complex architecture of a biological cell is of course not adequately captured by a liquid droplet and many other mechanisms are known to contribute to transport in a cell. 
Nevertheless such a simple model as a self-propelled droplet can be a first approximation to study the chaotic flow generated by active matter in  
biological cells. It may furthermore prove useful for efforts to synthesize cells from a small number of components.

\section{Chaotic advection from internal self-propelling flow}
In the present work, we consider autonomous self-propelled droplets,
driven by an internal flow. We show that the trajectories of particles
advected by this flow display the full richness of dynamical systems,
ranging from stagnation points and closed orbits to quasi-periodic
motion and chaos~\cite{Moffatt1990,Aref_2017}. Self-generated
flow provides an efficient mechanism for stirring and mixing inside
the droplet, while the motion of the droplet as a whole remains simple
and regular, e.g.  rectilinear or along a helix.

We study  spherical droplets of radius $a=1$, built
from two immiscible fluids and driven either by active body force
distributions, $\mb{f}^{act}$ inside the droplet or by active surface
tractions $\mb{t}^{act}$ on the interface. 
The forcing
mechanism generates flow $\vb(\rb)$ in the interior of the droplet,
which in turn generates flow in the surrounding fluid and gives rise
to self-propulsion.  We have computed such flow fields for a spherical
droplet and general forcing in Stokes approximation~\cite{Kree_2017},
i.e. the solution of
\begin{equation}
	\eta\nabla^2 \mb{v}-\nabla p=-\mb{f}^{act}, \quad \nabla\cdot \mb{v}=0.
\label{eq:stokes}
\end{equation}
If the forcing is achieved by tractions, then $\mb{f}^{act}=0$
and force balance at the interface has to include active tractions $\mb{t}^{act}$. Self-propulsion requires that the total force and torque
exerted by the fluid on the droplet vanishes.

The internal flow implies a constant nonzero linear and angular momentum of the droplet
which determine its linear and angular velocity,
$
  {\bf U}=\frac{3}{4\pi}\int_V\mb{v}(\mb{r}) d^3 r$ and $ \boldsymbol{\omega}
   =\frac{15}{8\pi}\int_V d^3 r \,\mb{r} \times  \mb{v},
$
of self-propulsion. 
The simplest such flow field due to active tractions, which propels the droplet  in $\hat{\mb{n}}$-direction  takes on the form    
 \begin{align}
  {\mb v}_{t}({\bf r})=  (1-2r^2)\hat{\mb{n}}+ (\hat{\mb{n}}\cdot\mb{r})\mb{r}
\label{eq:vminus}
\end{align}
in the co-moving frame. A force- and torque-free rotational flow  
can be written as
\begin{equation}\label{eq:rotationalflow}
  {\bf v}_{r}({\bf r})= \,h(r){\mb r}\times {\hat{\boldsymbol{\omega}}}.
\end{equation}
with the unit vector $ \hat{\boldsymbol{\omega}}= \boldsymbol{\omega}/|\boldsymbol{\omega}|$ pointing in the direction of angular momentum. 
Such a flow cannot be generated by tractions but needs chiral body forces.  For details see  \cite{Kree_2017} , where we also derived admissible functional forms of $h(r)$. All these flows must vanish on the interface. For the present work, we have chosen $h(r)=-1+3r^2-2r^3$ as a simple illustrating example. 
Advected particles subjected to a linear combination of $\vb_{t}$ and $\vb_{r}$  follow the equation of motion
$
  \frac{d{\rb}}{dt}=a_{t}\vb_{t}(\rb) + a_{r}\vb_{r}(\rb),
$
which constitutes a   three
dimensional, non-linear dynamical system in the interior of a sphere. 
In the following we first
introduce axially symmetric regular flow fields characterised by $\hat{\mb{n}}=\hat{\boldsymbol{\omega}}$. This idealised situation is perturbed either by a time-independent biaxiality ($\hat{\mb{n}}\neq \hat{\boldsymbol{\omega}}$) or by a time-periodic  direction $\hat{\mb{n}}(t)$. 
We discuss that both perturbations, taken separately or in combination,
give rise to chaotic trajectories and to mixing.

For  $\hat{\mb{n}}=\hat{\boldsymbol{\omega}}=\hat{\mb{z}}$ the equations of motion 
in cylindrical coordinates $x=\rho\cos\phi,\, y=\rho\sin\phi, \rho=\sqrt{x^{2}+y^{2}} $  take on the form
\begin{align}
  \frac{d\rho}{dt}=& a_tz\rho\nonumber\\
  \frac{d z}{dt}=&a_t(1-2\rho^2-z^2)\nonumber\\
 \frac{d\phi}{dt}=&-a_{r} h(\rho, z),
\end{align}
 which implies the conservation of angular momentum $L_z$ due to axial symmetry.  
 As for any two-dimensional incompressible flow, the first two equations can be mapped onto Hamiltonian
dynamics.  For our case the Hamiltonian  becomes
$H=a_t(p^2+pq^2-p)$ with $q=z,p=\rho^2$. 

\begin{figure}
 \stackon{(a)}{ \includegraphics[width=0.41\columnwidth]{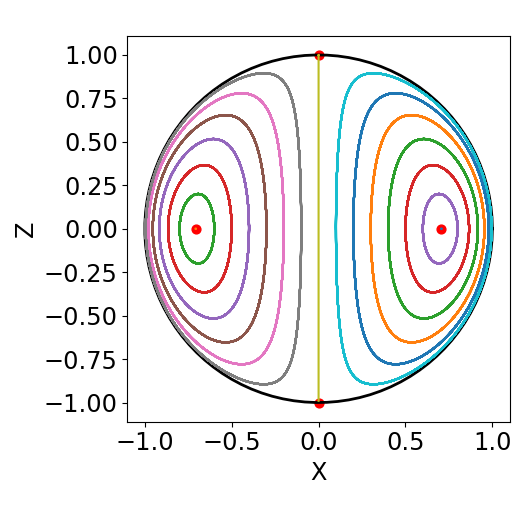}}
   \stackon{(b)}{ \includegraphics[height=0.42\columnwidth]{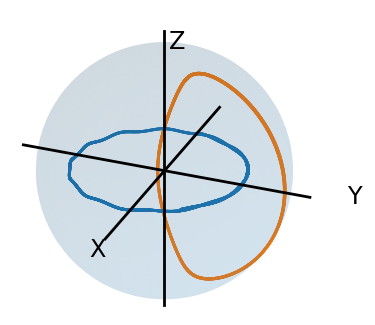}}
\caption{\label{fig1}  (a): Trajectories of tracer particle inside a  droplet, which moves with constant linear velocity and without rotation. Initial conditions are $\mb{r}(0)=\pm k\cdot 10^{-1}\,\hat{\mb{x}}$ with $k=1,\cdots 6$ (b): 
a 3d graph of the cycles for $\hat{\boldsymbol{\omega}}$ tilted with respect to $\hat{\mb{z}}$ by $\theta=0.3\pi$ (see also Fig.\ref{fig2})} 
\end{figure}

For $a_r=0$ (and thus $L_{z}=0)$ the trajectories in the x-z plane are shown in the left part of Fig.\ref{fig1}. Also shown are the hyperbolic stagnation points at  the poles, and the two elliptic stagnation points at $z=0,\, \rho=\pm1/\sqrt{2}$, which  span a whole circle of stagnation points in the x-y plane due to the axial symmetry. For any other choice of $\hat{\mathbf{n}}=\hat{\boldsymbol{\omega}}$, the flow pattern is a rotated version  of   Fig.\ref{fig1}a.
If the strength of the forcing and thus  $a_t(t)$ is time-dependent, this can  be absorbed into a
rescaled time $\tau$ with $d\tau/dt=1/a_t(t)$ without changing the streamlines and the shape of the droplet's trajectory.  In the following, we put 
$a_t=1$. 

For $a_r\neq 0$ the angular momentum
$L_z$ is still conserved, but no longer vanishes. Every trajectory results from  
 a superposition of two periodic motions, one in the x-z plane due to $\mb{v}_{t}$ and
one in the x-y plane due to $\mb{v}_{r}$, which leads to  periodic or quasi-periodic motion on  tori.  
It is plausible and can be shown that these tori constitute a complete foliation of the  sphere.
 
\begin{figure}[!tb]
 \stackon{(a)}{ \includegraphics[width=0.45\columnwidth]{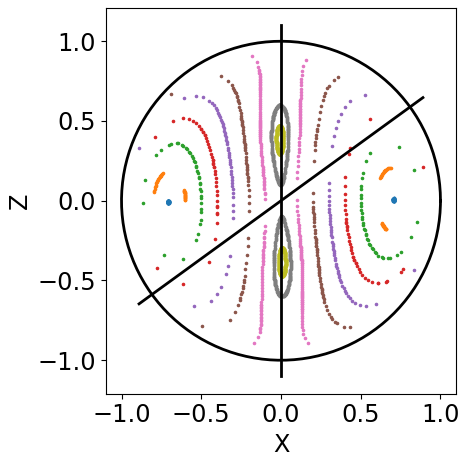}}
  \stackon{(b)}{  \includegraphics[width=0.45\columnwidth]{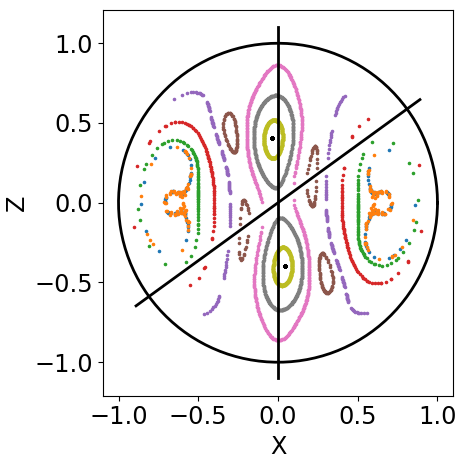}}\\
     \stackon{(c)}{ \includegraphics[width=0.45\columnwidth]{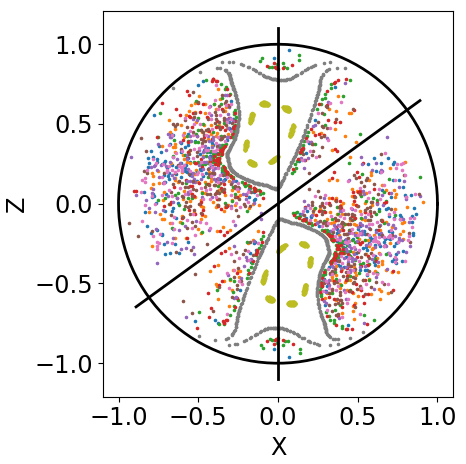}}
     \stackon{(d)}{ \includegraphics[width=0.45\columnwidth]{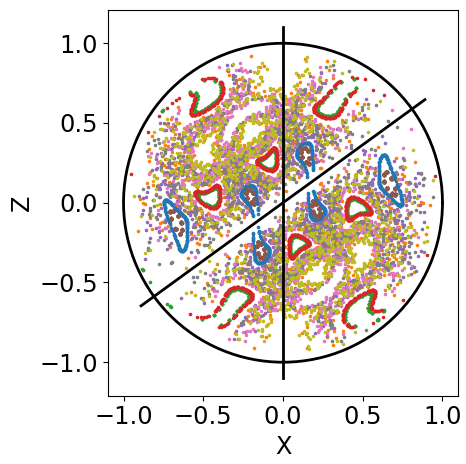}}

  \caption{\label{fig2} Poincar\'{e} section (x-z plane) of time-independent flow with $\hat{\mathbf{n}}=\hat{\mathbf{z}}$ and 
 $\hat{\boldsymbol{\omega}}$ tilted as indicated in the figure for different strength of the rotational flow, 
  $a_{r}=$ $1.1$ (a),  $2$ (b),  $3$ (c),  $9$ (d); initial conditions as used in Fig.\ref{fig1}a, with 2 additional trajectories starting on the z-axis at  $0.1\cdot \hat{\mb{z}}$ and $0.3\cdot \hat{\mb{z}}$}.  
\end{figure}

Now consider trajectories of advected particles, which are generated
by time-independent forcing for which the translational and angular
velocities are {\it no longer parallel}. As a consequence, the system is not
axially symmetric and the trajectories are determined by a fully
coupled, autonomous system of three differential equations.  There are
two control parameters of this flow: $a_r$ and
$\hat{\mathbf{n}}\cdot\hat{\boldsymbol{\omega}}=\cos\theta$.  To
illustrate the transition to chaos, we choose
$\hat{\mb{n}}=\hat{\mb{z}}$ and $\hat{{\boldsymbol{\omega}}}$ tilted
by an angle $\theta$ with respect to the z-axis in the x-z plane.  In
Fig.\ref{fig2}, we show Poincar\'{e} sections of this flow for
$\theta=0.3\pi$ and different $a_{r}$.  For small values of $a_{r}$,
we find regular motion for all values of $\theta$.
A finite tilt of the $\hat{\boldsymbol{\omega}}$ axis causes a new
cycle to appear, corresponding to 2-periodic points in the
Poincar\'{e} section, see Fig.\ref{fig2}. Now there are two types
of tori (A and B), winding around the two cycles shown in Fig.\ref{fig1} b.
Type A wind around the cycle, which continuously emerges
from the line of fixpoints at $a_r=0$, and type B tori  enclose 
the new  cycle. 
 The topology of the flow is controlled by these
coherent structures and the vanishing of  the rotational velocity and
the radial part of the translational velocity on the
interface. 
This latter feature is a general property of \emph{all} force- and
torque-free flows inside the sphere.
Thus advected particles on the surface always follow
regular trajectories connecting the two poles. In
terms of dynamical system theory, the droplet's surface is the
unstable manifold of the south pole and the stable manifold of the
north pole, which are hyperbolic fixpoints. 
The rotational flow also vanishes on the axis of
rotation, which therefore is clearly visible in the Poicar\'{e}
sections. 
With increasing $a_{r}$ tori of
both types decay and chaotic trajectories as well as islands of
regular motion appear.
We stress that the motion of the droplet as a whole is
still simple. If we assume that the time-independent forcing is
co-translating and co-rotating with the interior fluid, then the
droplet moves on a helix -- even in the presence of chaotic
streamlines inside. (For a detailed discussion of the droplet's
trajectories in dependence on the internal flow see
Ref.\cite{Kree_2017}).

\begin{figure}[!tb]
 \stackon{(a)}{ \includegraphics[width=0.45\columnwidth]{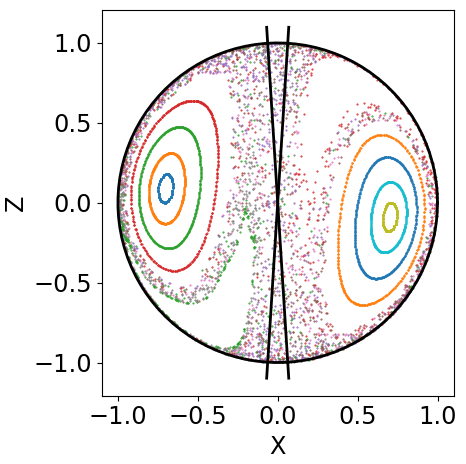}}
   \stackon{(b)}{\includegraphics[width=0.45\columnwidth]{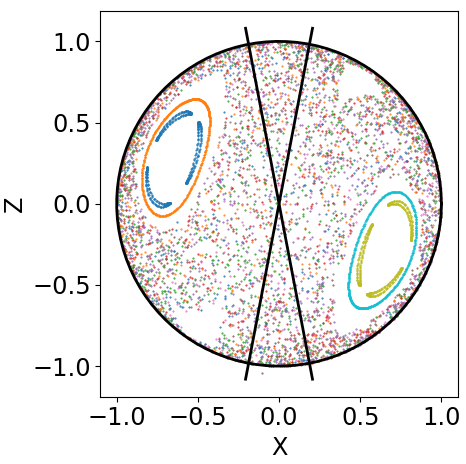}}\\
     \stackon{(c)}{\includegraphics[width=0.45\columnwidth]{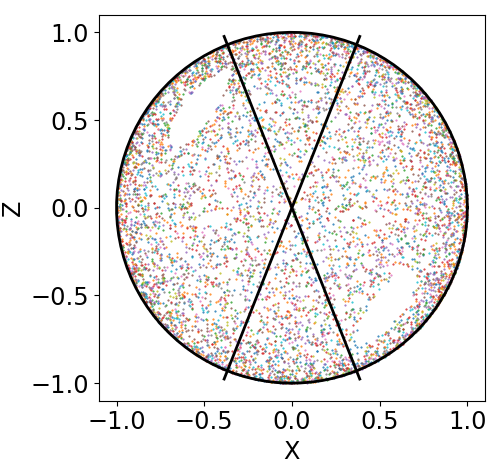}}
      \stackon{(d)}{ \includegraphics[width=0.45\columnwidth]{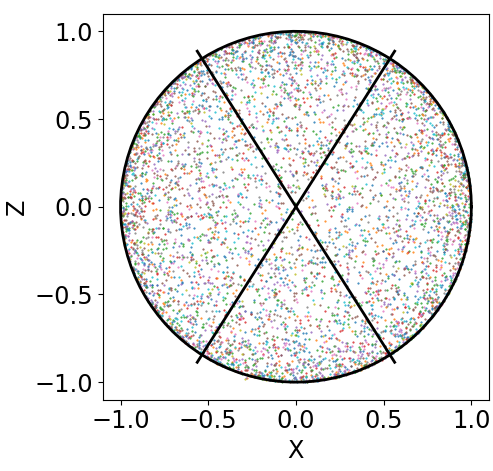} }
       \caption{\label{fig3} Stroboscopic views  for a flow due to time-dependent $\hat{\mathbf{n}}$ (see text); tilt angles $\Delta \theta =k\pi/180$ with k=4 (a), 10 (b), 20 (c) and  30 (d) for same initial conditions as in Fig.\ref{fig2}a }
\end{figure}

In biological systems the active elements such as motors in a vesicle or cell, are in general time dependent.
Of particular interest is a time-periodic forcing. Here, we consider a droplet without rotational flow but with 
periodic changes in the droplet's swimming direction, $\hat{\mb{n}}(t) =\hat{\mb{r}}(\theta(t), \varphi(t))$,   parametrised by  
polar and azimuthal angles $\theta(t)$ and $\varphi(t)$. If we choose 
$\varphi(t)=0$ the $\hat{\mb{n}}$-axis aways stays in the x-z plane. 
 It is easy to see by direct inspection of the equations of motion that trajectories starting with $y=0$  remain in the x-z plane, which  thus constitutes a 2d invariant manifold of the 3d flow. 
Consequently, the 3d flow originating from initial conditions with $y > 0$ ($y<0$) is restricted to the corresponding half-sphere.  
In the following we consider a simple harmonic time dependence
$\theta(t)=\Delta\theta \cos(t)$ and  the same initial conditions ($y=0$) as those shown in  Fig.\ref{fig1}a.
Stroboscopic views (at $t_k=2\pi k$) of trajectories are shown in  Fig.\ref{fig3}  for increasing amplitude $\Delta\theta$.   
For small $\Delta\theta$ we observe coexistence of regular motion and chaotic trajectories, but regular orbits gradually disappear for larger tilt. 
The flow within the 2d invariant manifold represents an example of the classical scenario of emerging chaotic behavior in a periodically driven Hamiltonian system in a 2d phase space\cite{Arnold1989}.
The onset of chaos does not affect the motion of the droplet as a whole, which is a simple translational motion following the oscillation of $\hat{\mb{n}}(t)$. 

\section{Mixing properties}
\begin{figure}[!tb]
  \stackon{(a)}{\includegraphics[width=0.4\columnwidth]{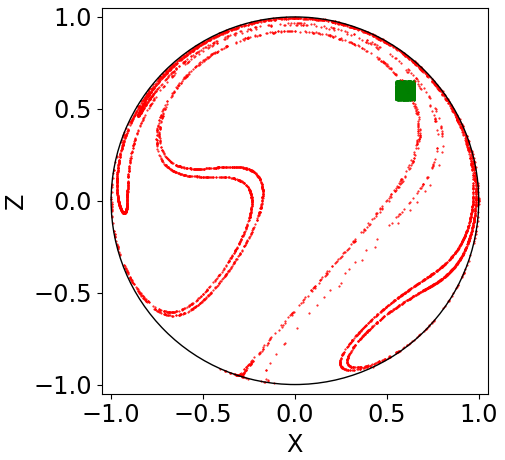}}
  \stackon{(b)}{\includegraphics[width=0.4\columnwidth]{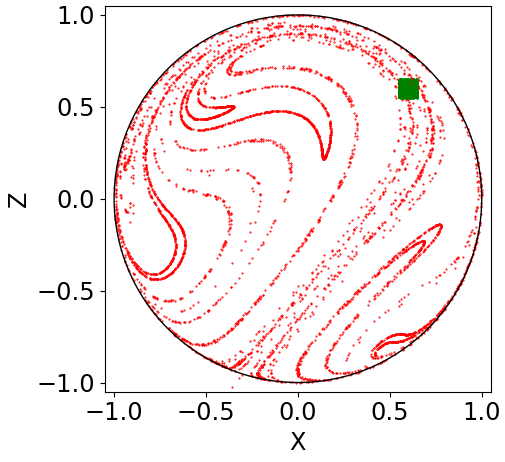}}\\
   \stackon{(c)}{ \includegraphics[width=0.4\columnwidth]{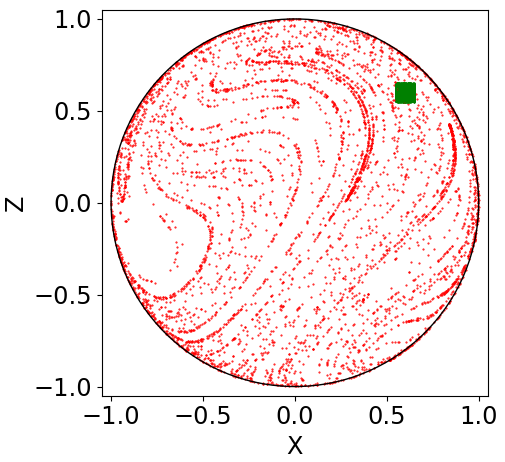}}
     \stackon{(d)}{ \includegraphics[width=0.4\columnwidth]{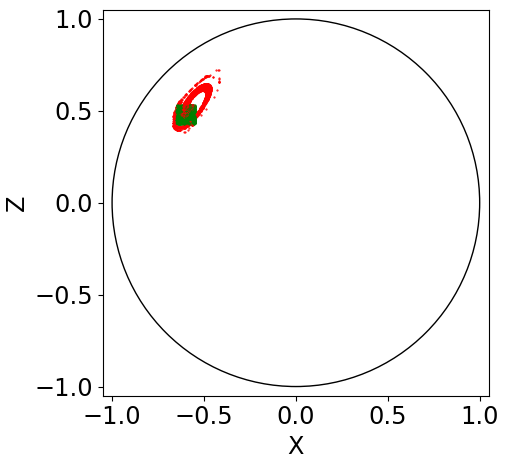}}
     \caption{\label{fig4} Temporal development of an initial
       distribution of $5\cdot 10^3$ particles distributed at random
       within the solid square centered at $\mb{r}=(0.6, 0, 0.6)$ for
       3 (a), 4 (b) and 5 (c) periods of $\theta(t)$, with
       $\Delta\theta=0.2\pi$.
       d) 
       centered at $\mb{r}= (-0.6, 0, 0.5)$ for $\Delta\theta=0.12$
       after 30 periods of $\theta(t)$, }
\end{figure}

The chaotic trajectories provide an efficient internal mechanism for
mixing, which may accelerate intracellular processes like signal
transduction, as we now discuss.  There are mathematical definitions
of perfect mixing in ergodic theory \cite{Arnold1989} , but there is
no unique measure of the efficiency of mixing. Here we focus on a simple
generic scenario as it may arise in signal transduction,--- the
temporal development of a chemical signal of $N$ initially
neighbouring particles. An example is shown in Fig.\ref{fig4} for
the 2-dimensional flow described above.  Mixing efficiency is
characterised by the dispersal of the initial signal for times up to a finite 
$t_{max}$, which is set by reaction times, life times of
active states or other relevant scales.  Stretching, folding and
stirring is capable to transport a chemical signal over the droplet‘s
diameter within a few periods of oscillation, while at the same time
it leads to a distribution of large concentration gradients, as seen
in Fig.\ref{fig4}a. Once several such stretching and folding
processes have taken place (see Fig.\ref{fig4}b), diffusive
transport is expected to take over and generate a homogeneous
distribution in a cell or vesicle of size in the $\mu m$ range (see
below.) Note, however that not all trajectories are chaotic in
Fig.\ref{fig3}. We also observe regular islands, and if the initial
distribution is chosen within such an island it does not mix, as
illustrated in Fig.\ref{fig4}d. Regions of chaotic advection are thus
separated from regular motion by transport barriers, which can only be
surmounted by diffusion. In this way the droplet’s interior is
subdivided dynamically into mixing and non-mixing regions, which can
be controlled via parameters of the flow. An active droplet can
thus easily switch from a mixing to a  partially mixing or non-mixing
state, by changing its global activities, like e.g.,  $\Delta\theta$ or the velocity
scales $a_{r}$ and $a_{t}$ of chiral and non-chiral flow.

\begin{figure}[!tb]
     \stackon{(a)}{\includegraphics[width=0.45\columnwidth]{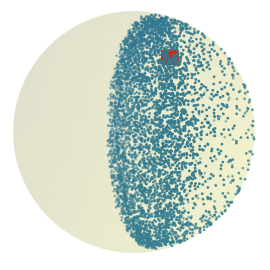}}
      \stackon{(b)}{ \includegraphics[width=0.45\columnwidth]{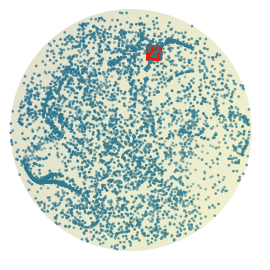} }
       \caption{\label{fig5} Positions of $5\times 10^{3}$ particles initially distributed randomly within a small solid cube centered at  $\mb{r}_{0}=(0.6, 0.1, 0.5)$ after 6 periods of  oscillations of $\hat{\mb{n}}(t)$ with    $\Delta\theta=0.2\pi$ as in Fig. \ref{fig4}a-c. (a) without rotational flow and (b) with rotational flow of strength $a_{r}=1$ }
\end{figure}

So far we have discussed mixing in the x-z plane, which constitutes an invariant manifold of the flow. Mixing in 3d can be achieved by either choosing the initial conditions outside the invariant manifold or by adding a rotational flow. 
In the former case, the invariant manifold  constitutes a barrier between the $y>0$- and the $y<0$-half-sphere.  Fig. \ref{fig5}a is the analogue of
 Fig. \ref{fig4}c for an initial distribution in a cube centered at $\mb{r}_{0}=(0.6, 0.1, 0.5)$. The resulting particle distribution fills the $y > 0$ half-sphere, however, it is not uniform but accumulates in the neighborhood of the invariant manifold.  Mixing within the entire sphere can be achieved with an additional rotational flow as shown in Fig. \ref{fig5}b. Here the rotational flow of Eq.{\ref{eq:rotationalflow}} has been added with $\hat{\boldsymbol{\omega}}=\mb{e}_{z}$ and $a_{r}=1$. 
This rotational flow transports the 2d mixing in the invariant manifold into the sphere, and thus leads to effective mixing in 3d. 
 
From the particle distribution  we can derive further quantitative statistical measures of mixing by coarse graining.    
We introduce a partitioning of 2d or 3d space  into $M$ cells with areas or volumes $\{F_n\}_{n=1}^{M}$. In each cell, we count the number of advected particles $N_n$ and measure the density $\rho_n=N_n/(F_n N)$, which is an estimate of the probability to find a particle in the cell. In order to compare this probability to a uniform distribution, we use the Kullback-Leibler (KL) entropy, defined as
\begin{equation}
S=\sum_n \rho_n \log\frac{\rho_n}{\rho}  
\end{equation}
with the uniform density $\rho=1/F$. Small values of $S$ indicate good homogeneous mixing properties, whereas large values result from strongly inhomogeneous distributions as, e.g., in Fig.\ref{fig4}d. 

\begin{figure}[!tb]
\def\big{ \includegraphics[width=\columnwidth]{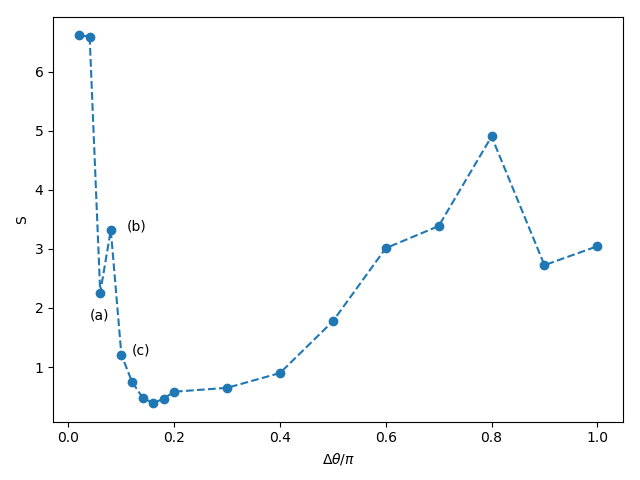}}
\def\littlea{ \stackon{(a)}{ \includegraphics[width=0.15\columnwidth]{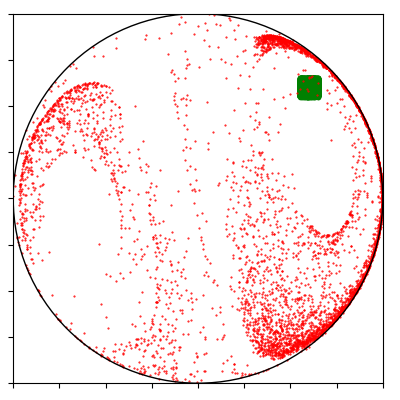}}}
\def\littleb{ \stackon{(b)}{ \includegraphics[width=0.15\columnwidth]{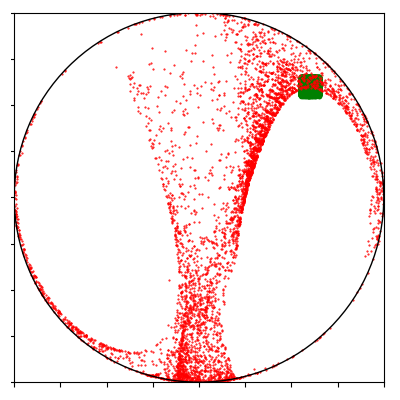}}}
\def\littlec{ \stackon{(c)}{ \includegraphics[width=0.15\columnwidth]{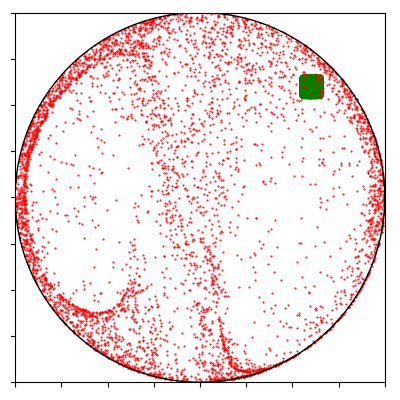}}}
\savestack{\firststack}{\stackinset{l}{45pt}{t}{23pt}{\littlea}{\big}}
\savestack{\secondstack}{\stackinset{l}{95pt}{t}{23pt}{\littleb}{\firststack}}
\stackinset{l}{145pt}{t}{23pt}{\littlec}{\secondstack}
\caption{\label{fig6} Coarse grained Kullback-Leibler entropy at $t=24\pi$ versus $\Delta\theta$ for 
  the initial distribution shown in Fig.\ref{fig4}a; 
  insets: temporal development as in Fig.\ref{fig4} for 3 selected values (a),(b),(c) of $\Delta \theta$ and 12 periods of $\theta(t)$.
}
 \end{figure}

The KL-entropy depends upon bin sizes and contains statistical errors due to finite N. We have studied these dependencies and found that $N \geq 5\cdot 10^{3}$ particles and $\geq 25$ bins in both angular and radial coordinates  are sufficient to detect global features of mixing. 
 The entropy is plotted versus
$\Delta\theta$ in Fig.\ref{fig6}  after 12 cycles of the axis $\mb{n}$, which we found to be sufficient to reach a stationary value.
The initial conditions have been chosen  as in Fig.\ref{fig4}a. 
For small values of  $\Delta\theta \lesssim 0.2$  one observes a decrease of entropy. 
The decrease is, however, 
non-monotonic due to Lagrangian coherent structures (like islands, for example), which appear and vanish as $\Delta\theta$ is varied. 
This is shown for 3 examples in the inset of Fig.\ref{fig6} a-c. 
Thus homogenization of an initially localized distribution of particles depends sensitively on $\Delta\theta$ in accordance with our previous results on single particle trajectories. The coexistence of chaotic regions and regular islands is reflected in correspondingly complex homogenization and mixing properties. 


\section{Conclusions and Discussion}
To conclude, flow inside a self-propelling droplet driven by internal active forcing exhibits chaotic advection,
which should be observable in an experiment.
We stress that the chaotic behavior is predicted for autonomous swimmers, whose trajectories are simple and regular. Although we have illustrated the emergence of chaotic trajectories and their effects on mixing only for a special, self-propelling flow,  it is known that there are generic routes to chaos in 3d incompressible flows~\cite{Mezic1994}, and thus we expect most of our qualitative conclusions to hold in a much broader context.

We have investigated the mixing properties of the internal flow by following the time development of an initially localised set of particles, as well as with the help of the Kullback-Leibler entropy. Mixing in microfluidic devices has been studied extensively~\cite{Lee2011}. Given the reversibility of the flow, mixing is usually achieved with the help of carefully designed geometries or time-dependent external forces. Here, in contrast, we consider spherical shapes only and show that internal forcing mechanisms 
result in mixing. 
How relevant are these results on scales of biological cells where molecules or vesicles need to be transported? Can chaotic advection beat  transport or mixing by diffusion? As a hint, we estimate two dimensionless numbers for an internal fluid with a viscosity $\eta$. For directed transport, the P\'{e}clet number
$
  Pe=UL/D
$
gives the ratio of convective to diffusive transport.  
Here, $U$ denotes a typical velocity, $L$ the size of the cell, and for an object of radius $R$ we estimate 
the diffusion constant  $D=k_{B}T/6\pi\eta R$ with help of  the Einstein relation.  
Assuming $\eta$ to be  4 times the viscosity of the ambient water
we find $Pe/(ULR)\approx 6\pi\, s/\mu m^{3}$.  The extension of a diffusing protein is estimated as $R\sim10\, nm$. For a bacterium of typical cell size $L\sim  1\, \mu m$, one needs a velocity of   
$U\sim 5\, \mu m/s$ to achieve $Pe = 1$. On the other hand, a eukaryotic cell can be as large as $L \sim 100\, \mu m$, implying $Pe\sim 100$ for the same velocity.  A typical scale, which compares mixing by stretching and folding  to diffusion is the Batchelor length $l_{B}=\sqrt{D/\lambda}$. For distances smaller than $l_{B}$ mixing is dominated by diffusion~\cite{Aref_2017}. Here $\lambda$ denotes a positive Liapunov exponent, which characterizes the spreading of two nearby initial conditions: $\delta x \sim e^{\lambda t}$. We found that $\lambda \sim 1/s$ for the studied chaotic flows with typical velocities of $1\, \mu m/s$ so that $l_{B} \sim 1\, \mu m$. Thus chaotic advective transport in an autonomous microswimmer provides an efficient mechanism for mixing in case diffusion is slow due to the large size  of either  the advected object or the cell. Even for small sized objects or cells, the rapid buildup of large gradients due to stretching and folding can accelerate diffusive transport substantially.

\section*{Appendix}

In this apendix, we briefly recall the computation of the flow fields
inside the droplet, and in particular the rotational flow of
Eq.\ref{eq:rotationalflow}, which cannot be generated by surface tractions but requires
the action of body forces. For details of the calculations we refer to
\cite{Kree_2017}.  

Consider a
neutrally buoyant spherical droplet of radius $R=1$, filled with an
incompressible Newtonian fluid of viscosity $\eta$, which is swimming
in another Newtonian fluid of viscosity $\eta_{a}$ and driven by
internal active volume force densities $\mb{f}(r, \theta, \varphi)$
and/or active surface force densities $\mb{t}( \theta,\varphi)$ on the
interface at $r=R$. The
forces generate a flow field inside the droplet, which is
coupled to the ambient fluid via viscous forces at the interface and
may lead to self-propulsion. To calculate the flow field $\mb{v}$ in
the lab frame we choose a spherical coordinate system
$(r, \theta, \varphi)$ with its origin at the center of the droplet at
a fixed time $t$.  The unit vectors corresponding to the coordinate
lines are denoted by $\mb{e}_{r}, \mb{e}_{\theta}, \mb{e}_{\varphi}$.

For small Reynolds number, the flow obeys Stokes' equations
\begin{equation}
\eta(r) \nabla^{2} \mb{v} =-\nabla p +\mb{f} ; \quad \nabla\cdot \mb{v}=0
\end{equation} 
with $\eta(r<1)=\eta$ and $\eta(r>1)=\eta_{a}$.  To distinguish the
internal from the ambient flow we introduce $\mb{v}(r>1)=\mb{v}_{a}$.
For immiscible fluids and no interfacial velocity slip, $\mb{v}$ is
continuous across the interface,
\begin{equation}
\mb{v}(r=1, \theta, \varphi)= \mb{v}_{a}(r=1, \theta, \varphi).
\end{equation} 
Furthermore every surface element of the interface is force free, which implies 
\begin{equation} 
 \quad (\boldsymbol{\sigma}(\mb{v}_{a})-\boldsymbol{\sigma}(\mb{v}) - \mb{t})\cdot\mb{e}_{r}=2 \gamma\mb{e}_{r}.
 \label{eq:boundaryforcefree}
 \end{equation}
 The viscous stress tensor $\boldsymbol{\sigma}$ is characterized by
 its cartesian components
 $\sigma_{ij}(\mb{v})=-p\delta_{ij}+\eta(\partial_{i}v_{j}+\partial_{j}v_{i})$
 with pressure field
 $p(r,\theta,\varphi)$. 
 The term on the r.h.s. of Eq. \ref{eq:boundaryforcefree} results
 from a homogeneous surface tension $\gamma$.

To take advantage of the spherical geometry, we expand the pressure field into spherical harmonics $Y_{{\ell m}}$ and the vector fields $\mb{v}, \mb{f}$ and $\mb{t}$ into vector spherical harmonics. Our choice for the latter is $\mb{Y}^{(0)}_{\ell m}= Y_{\ell m} \mb{e}_{r}$, $\mb{Y}^{(1)}_{\ell m}= r\nabla Y_{\ell m}$ and  $\mb{Y}^{(2)}_{\ell m}= \mb{r}\times\nabla Y_{\ell m}$. Inserting the expansion
\begin{equation}\label{eq1}
\mb{v}(r,\theta,\varphi)=\sum_{s=0}^{2}\sum_{\ell=0}^{\infty}\sum_{m=-\ell}^{\ell} v^{s}_{\ell m}(r)\mb{Y}^{(s)}_{\ell m}(\theta, \varphi)
\end{equation}
and the corresponding expansions for $p, \mb{f}$ and $\mb{t}$ into Stokes equations results in a system of ordinary differential equations for 
$v^{s}_{\ell m}(r)$, which is completely decoupled in $\ell$ and $m$. For each $\ell, m$ the only remaining coupling is between $s=0$ and $s=1$ modes.    

The simple translational flow of Eq.\ref{eq:vminus} represents the
$m=0$ component of the nonchiral part of Eq.\ref{eq1} $(s=0,1)$
\begin{align}
  \mb{v}_t(\mb{r})&= \mb{v}(\mb{r})-\mb{U}\nonumber\\
  &\propto (r^2-1)\mb{Y}^{(0)}_{1 0}(\theta, \varphi)
  +(2r^2-1)\mb{Y}^{(1)}_{1 0}(\theta, \varphi)\nonumber\\
&\propto(r^2-1)\cos{\theta}\,\mb{e}_{r}+(1-2r^2)\sin{\theta}\,\mb{e}_{\theta}\nonumber
\end{align}
in agreement with Eq.\ref{eq:vminus}.
The rotational flow represents the chiral part of
Eq.\ref{eq1} $(s=2)$. Traction forces give rise to rigid body motion which is
not compatibel with the vanishing of the total torque as required for
autonomous swimming. Body forces on the other hand can give rise to
rotational flow inside the droplet. The simple example, discussed in
the main text, corresponds to the force density
$\mb{f}(\mb{r})=\sum_m\gamma_m(r)\mb{Y}^{(2)}_{1 m}(\theta, \varphi)$.
Let us first consider the axially symmetric case with only the $m=0$ component
of the force density.
The total torque vanishes, if $\int_0^1 dr r^3\gamma_0(r)=0$. This can be achievd by two power laws for $\gamma_0(r)$, for example $\gamma_0(r)=c_1r+c_2r^2$
with
$c_1/5=-c_2/6$.
The resulting rotational flow is given by 
\begin{align}
  \mb{v}_r(\mb{r})&\sim\big((r-r^3)/2+(r-r^4)/3\big)
                                \mb{r}\times\nabla Y_{1 0}(\theta,\varphi)\\
                              &\sim h(r)\sin{\theta}\,\mb{e}_{\phi}=
                                -h(r)\mb{r\times}\mb{e}_{z}.
                                \nonumber
\end{align}
The more general case with all $\gamma_m(r)$ of the same functional form,
can then be shown~\cite{Kree_2017} to result in
$ \mb{v}_r(\mb{r}) \sim -h(r)\mb{r\times}{\hat{\boldsymbol{\omega}}}$.

\end{document}